\documentclass[preprint,11pt]{elsarticle}

\usepackage{graphicx}
\usepackage{amsmath}
\usepackage{amssymb}
\usepackage{hyperref}
\usepackage{dcolumn}
\usepackage{bm}
\usepackage{color}
\usepackage{textcomp,appendix}

\journal{Annals of Physics}

\begin{document}

\begin{frontmatter}

\title{Life-times of quantum resonances through the Geometrical Phase Propagator Approach}

\author{G.~Pavlou}


\address{Department of Physics, University of Athens, GR-15771 Athens, Greece}

\author{A.~I.~Karanikas}


\address{Department of Physics, University of Athens, GR-15771 Athens, Greece}

\author{F.~K.~Diakonos}


\address{Department of Physics, University of Athens, GR-15771 Athens, Greece}

\date{\today}

\begin{abstract}

We employ the recently introduced Geometric Phase Propagator Approach (GPPA)\cite{Diakonos2012} to develop an improved perturbative scheme for the calculation of life times in driven quantum systems. This incorporates a resummation of the contributions of virtual processes starting and ending at the same state in the considered time interval. The proposed procedure allows for a strict determination of the conditions leading to finite life times in a general driven quantum system by isolating the resummed terms in the perturbative expansion contributing to their generation. To illustrate how the derived conditions apply in practice, we consider the effect of driving in a system with purely discrete energy spectrum, as well as in a system for which the eigenvalue spectrum contains a continuous part. We show that in the first case, when the driving contains a dense set of frequencies acting as a noise to the system, the corresponding bound states acquire a finite life time. When the energy spectrum contains also a continuum set of eigenvalues then the bound states, due to the driving, couple to the continuum and become quasi-bound resonances. The benchmark of this change is the appearance of a Fano-type peak in the associated transmission profile. In both cases the corresponding life-time can be efficiently estimated within the reformulated GPPA approach.

\end{abstract}

\begin{keyword}
quantum resonances \sep life-time \sep perturbation theory \sep resummation


\end{keyword}

\end{frontmatter}

\section{Introduction}
\label{intro}

In addition to that driven quantum systems are often associated with quantum resonances \cite{Lax1967}, an active field for experimental and theoretical studies related to a variety of interesting phenomena. Some of them include the controlled emergence or the coherent destruction of stochastic resonances in tunneling processes \cite{Gammaitoni1998}, the resonant energy transfer in quantum dot semiconductors with impact on medical imaging \cite{Willard2003} and the resonant conduction in electronic nano-devices \cite{Bulka2001}. Open quantum systems constitute a large class supporting resonant states. In this class belong also the driven systems. The driven systems play a crucial role in the study of quantum resonances since the involvement of an external field enables in principle the outer influence to the induced resonant behaviour \cite{Mabuchi2004}. This possibility along with recent technological developments in the design of external laser or magnetic fields of arbitrary profile open a new area in quantum physics targeting the manipulation of quantum resonant behaviour \cite{Rabitz2009}.

Despite their long history in quantum physics \cite{Lax1967} quantum resonances are still an active field for experimental and theoretical studies related to a variety of interesting phenomena especially at the mesoscopic scale. Typical examples of modern developments in the context of quantum resonances is the controlled emergence or the coherent destruction of stochastic resonances in tunneling processes \cite{Gammaitoni1998}, the resonant energy transfer in quantum dot semiconductors with impact on medical imaging \cite{Willard2003}, the resonant conduction in electronic nano-devices \cite{Bulka2001}, etc. Open quantum systems constitute a large class supporting resonant states. In this class belong also the driven systems which usually consist of a closed system coupled to an external time-dependent field. The driven systems play a crucial role in the study of quantum resonances since the involvement of an external field enables in principle the outer influence to the induced resonant behaviour \cite{Mabuchi2004}. This possibility along with recent technological developments in the design of external laser or magnetic fields of arbitrary profile open a new area in quantum physics targeting the manipulation of quantum resonant behaviour \cite{Rabitz2009}.  

Theoretically the description of resonances in the quantum regime is usually performed either by using the Feshbach formalism \cite{Feshbach1958} or the Gamow-Siegert boundary conditions for solving the corresponding Schr\"{o}dinger equation \cite{Siegert1939}. Also one can use complex scaling \cite{Aguilar1971} transforming the Gamow-Siegert eigenfunctions to square integrable ones in order to locate resonances as ordinary eigenvalues in the spectrum of the transformed Hamilton operator. These methods are proven to be equivalent for a wide range of applications and are usually employed to handle open quantum systems which do not depend explicitly on time. In the case of explicit time dependence and when the external driving is periodic, Floquet theory can be used to transform the original system to a more complicated stationary one \cite{Tannor2007}. When the external field is weakly coupled to the considered system the standard time-dependent perturbation theory to calculate life-times of quantum resonances applies \cite{Sakurai2014}. In a recent work an improved time-dependent perturbation scheme named Geometric Phase Propagator Approach (GPPA) has been proposed \cite{Diakonos2012}. It is founded on a reformulation of the standard (adiabatic) perturbation theory and it is valid for any kind (periodic or not) of driving. It has the ability of a dynamical description of quantum resonances and it allows simple physical interpretation of the calculated contribution to any related observable in terms of the number of transitions between states of a stationary spectrum describing the system at a given instant.

The aim of the present work is to show how the GPPA encapsulates the presence of quantum resonances, enabling with a suitable resummation to determine also the necessary conditions for the occurrence of states with finite life times. Furthermore a systematic pathway to calculate these life-times is given. To clearly demonstrate our procedure we will examine two examples. First we will consider the case of a driven bound system and we will show how the frequency spectrum of the driving field is related to the emergence of finite life times for the associated bound states which become quasi-bound due to the driving. Subsequently we will study the case of a system described by a Hamiltonian with mixed (discrete and continuous) eigenvalue spectrum. Based on the concept of the instantaneous stationary spectrum we will show that a monochromatic driving is sufficient to trigger transitions between the discrete and the continuous part of the spectrum transforming the states of the discrete part to quasi bound states with finite life time. The latter is calculated within the framework of GPPA utilizing the hierarchic classification of the contributing elementary transitions. Our treatment provides a unified scheme for the calculation of the quantum resonance properties and allows for remarkable insight concerning their origin. The paper is organized as follows. In Section \ref{arrange} we present a re-organized form of the adiabatic perturbation series based on the resummation of all the "loop" transitions that is, of all the transitions starting from and ending at the same state and we comment on whether this improved series expansion converges. In Section \ref{life} we apply the same resummation technique, which is the quintessence of GPPA, to calculate finite life-times and we express the resummed perturbation series in terms of the geometric phase propagators. In the same Section we take the opportunity to discuss the different physical scenarios leading to finite life-times. In Section \ref{ex} we apply the developed theoretical tool to specific physical examples. In \ref{dho} we calculate the life-time of quasi-bound states emerging in the spectrum of a harmonic oscillator driven by a polychromatic external field. We  emphasize on the role of the involved time scales due to the complex structure of the driving law. In \ref{delta} we apply the reformulated GPPA to calculate the life-time of a quasi-bound state in a harmonically driven delta-potential and we discuss the effect of this state on the associated transmission profile. Finally in Section \ref{end} we present our concluding remarks.

\section{Rearranging the adiabatic series}
\label{arrange}

In this section we briefly present the main ingredients of the GPPA introduced in \cite{Diakonos2012} using the opportunity to simplify significantly the formalism, increasing its friendliness for application to concrete physical systems. Furthermore, we perform a resummation of a class of terms in the perturbation expansion which turns out to be extremely adequate for calculating life-times. We begin by considering the dynamics of a quantum system described by the time-dependent Schr\"{o}dinger equation

\begin{equation}
\label{1.1.1}
{i\hbar {\partial _t}\left| {{\psi _t}} \right\rangle  = \hat H\left( t \right)\left| {{\psi _t}} \right\rangle ~~;~~\hat H\left( t \right) = {\hat H_0} + {\hat H_I}\left( t \right)}
\end{equation}

\noindent
assuming for the time dependent part of the Hamiltonian that: \\ ${\hat H_I}\left( { t \to \pm \infty } \right) = 0$. To solve Eq.~(\ref{1.1.1}) we expand as usual the state $\left| {{\psi _t}} \right\rangle $ in the  basis formed by the instantaneous eigenstates $\left| {{n_t}} \right\rangle $ of the time-dependent Hamiltonian

\begin{equation}
\label{1.1.2}
{\left| {{\psi _t}} \right\rangle  = \sum\limits_n {{\alpha _n}} \left( t \right)\left| {{n_t}} \right\rangle ~~;~~ \hat H\left( t \right)\left| {{n_t}} \right\rangle  = {E_n}\left( t \right)\left| {{n_t}} \right\rangle {\rm{ }}}
\end{equation}

\noindent
where ${E_n}\left( t \right)$ are the temporary energy eigenvalues. To simplify the notation we use only one index, the integer $n$, to label eigenstates. In general, however, the spectrum contains and a continuum part. Whenever necessary it will be explicitly denoted that the index belongs to the continuum. One can introduce renormalized temporary energy eigenvalues and transition matrix elements, subtracting suitably diagonal contributions:

\begin{equation}
\label{1.1.3}
{{\Phi _{nm}} = \left\langle {{n_t}} \right|i\hbar {\partial _t}\left| {{m_t}} \right\rangle  - \left\langle {{n_t}} \right|i\hbar {\partial _t}\left| {{n_t}} \right\rangle {\delta _{n,m}}~~;~~{\bar E_n}\left( t \right) = {E_n}\left( t \right) - \left\langle {{n_t}} \right|i\hbar {\partial _t}\left| {{n_t}} \right\rangle}.
\end{equation}

\noindent
Then it is possible to express the solution $\vert \psi_t \rangle$ in a more convenient form: 

\begin{equation}
\label{1.1.4}
{\left| {{\psi _t}} \right\rangle  = \sum\limits_{n,m} {{e^{ - \frac{i}{\hbar }\int\limits_{{t_i}}^t {d\tau {{\bar E}_n}\left( \tau  \right)} }}{X_{nm}}\left( {t,{t_i}} \right)} {\alpha _m}\left( {{t_i}} \right)\left| {{n_t}} \right\rangle}.
\end{equation}

\noindent
using a time-independent orthonormal basis $\{\vert n \rangle \}$ with $\sum\limits_n {\left| n \right\rangle } \left\langle n \right| = \hat 1$ and ${\alpha _n}\left( t \right) = \left\langle n \right|\left. {\alpha \left( t \right)} \right\rangle$. In Eq.~(\ref{1.1.4}) the transition matrix elements $X_{nm}\left( {t,{t_i}} \right)$ are given by:

\begin{equation}
\label{1.1.5}
{{X_{nm}}\left( {t,{t_i}} \right)={\left( {\hat T{e^{\frac{i}{\hbar }\int\limits_{{t_i}}^t {d\tau \hat \varphi \left( \tau  \right)} }}} \right)_{nm}}}
\end{equation}

\noindent
with:

\begin{equation}
\label{1.1.6}
{{\hat \varphi \left( t \right) = \sum\limits_{n,m} {\left| n \right\rangle } {\varphi _{nm}}\left( t \right)\left\langle m \right|}~~~;~~~{{\varphi _{nm}}\left( t \right) = {e^{\frac{i}{\hbar }\int\limits_{{t_i}}^t {d\tau {{\bar E}_n}} }}{\Phi _{nm}}\left( \tau  \right){e^{ - \frac{i}{\hbar }\int\limits_{{t_i}}^t {d\tau {{\bar E}_m}} }}}}.
\end{equation}

The completeness of the basis $\{ \vert n \rangle \}$ implies that $\sum\limits_n {{{\left| {{X_{nm}}} \right|}^2}}  = 1$. Without loss of generality we assume ${\alpha _m}\left( {{t_i}} \right) = {\delta _{m,0}}$ (where the index $"0"$ indicates the initial state, not necessarily the ground one) and split the above expression into two parts:

\begin{equation}
\label{1.1.7}
\begin{split}
\left| {{\psi _t}} \right\rangle & = \sum\limits_n {{e^{ - \frac{i}{\hbar }\int\limits_{{t_i}}^t {d\tau {{\bar E}_n}} }}{X_{n0}}\left( {t,{t_i}} \right)} \left| {{n_t}} \right\rangle \\ & = {e^{ - \frac{i}{\hbar }\int\limits_{{t_i}}^t {d\tau {{\bar E}_0}} }}{X_{00}}\left( {t,{t_i}} \right)\left| {{0_t}} \right\rangle  + \sum\limits_{n \ne 0} {{e^{ - \frac{i}{\hbar }\int\limits_{{t_i}}^t {d\tau {{\bar E}_n}} }}{X_{n0}}\left( {t,{t_i}} \right)} \left| {{n_t}} \right\rangle
\end{split}
\end{equation}

\noindent
 Leaving for the next section the discussion of the first term, we concentrate on the second one which contains the time-ordered exponential

\begin{equation}
\label{1.1.8}
{{X_{n0}} = {\left( {\hat T{e^{\frac{i}{\hbar }\int\limits_{{t_i}}^t {d\tau \hat \varphi \left( \tau  \right)} }}} \right)_{n0}} = \sum\limits_{r = 1}^\infty  {X_{n0}^{\left( r \right)}{\rm{  }}} ,~~~~n \ne 0}.
\end{equation}

\noindent
The matrix element ${X_{n0}^{\left( r \right)}{\rm{  }}}$ is written as:

\begin{equation}
\label{1.1.9}
\begin{split}
X_{n0}^{\left( r \right)} = & \sum\limits_{{n_1},...,{n_{r - 1}}} {{\left( {\frac{i}{\hbar }} \right)}^r}\int\limits_{{t_i}}^t {d{t_r}} \int\limits_{{t_i}}^{{t_r}} {d{t_{r - 1}}...\int\limits_{{t_i}}^{{t_2}} {d{t_1}} } {\varphi _{n{n_{r - 1}}}}\left( {{t_r}} \right)\times\\ & \times{\varphi _{{n_{r - 1}}{n_{r - 2}}}}\left( {{t_{r - 1}}} \right)...{\varphi _{{n_1}0}}\left( {{t_1}} \right)
\end{split}
\end{equation}

\noindent
In Appendix \ref{appen} it is shown in detail that it is possible to rearrange the terms in this series expansion obtaining a reformulation of Eqs.~(\ref{1.1.8},\ref{1.1.9}) with hierarchical ordering of sub-processes in terms of the number of transitions (flips) between different states of the basis $\{\vert n \rangle \}$. Among other advantages (which we do not discuss in this section) this reformulation allows for a resummation of classes of sub-processes leading to a direct estimation of life-times for a variety of physical set-ups. In practice the rearrangement of GPPA is performed introducing the matrix elements $X_{nn}^{\left[ {{n_1},..} \right]}\left( {{t_b},{t_a}} \right)$ which denote the probability amplitude a system starting the time $t_a$ from the state $n$, to end up to the same state at the time $ t_b$ without passing through the states ${n_1},{n_2},...$. Then Eq.~(\ref{1.1.8}) can be rewritten as:

\begin{equation}
\label{1.1.10}
{X_{n0}} = \sum\limits_{r = 1}^\infty  {X{{_{n0}^{\left( r \right)R}}}}
\end{equation}

\noindent
where

\begin{equation}
\label{1.1.11}
\begin{split}
X_{n0}^{\left( r \right)R} = & \sum\limits_{{n_{r - 1}} \ne ... = {n_1} \ne 0,n} {{{\left( {\frac{i}{\hbar }} \right)}^r}\int\limits_{{t_i}}^t {d{t_r}} \int\limits_{{t_i}}^{{t_r}} {d{t_{r - 1}}...\int\limits_{{t_i}}^{{t_2}} {d{t_1}} } X_{nn}^{\left[ {{n_{r - 1}},...,0} \right]}\left( {t,{t_r}} \right) \times} \\ & {\times {\varphi _{n{n_{r - 1}}}}\left( {{t_r}} \right)X_{{n_{r - 1}}{n_{r - 1}}}^{\left[ {{n_{r - 2}},...,0} \right]}\left( {{t_r},{t_{r - 1}}} \right)...X_{{n_1}{n_1}}^{\left[ 0 \right]}\left( {{t_2},{t_1}} \right){\varphi _{{n_1}0}}\left( {{t_1}} \right)\times} \\ & {\times{X_{00}}\left( {{t_1},{t_i}} \right)}.
\end{split}
\end{equation}

Each term in Eq.~(\ref{1.1.10}) attains a simple physical interpretation. For example the 
first order term:

\begin{equation}
\label{1.1.12}
{X_{n0}^{\left( 1 \right)R} = \frac{i}{\hbar }\int\limits_{{t_i}}^t {d{t_1}} X_{nn}^{\left[ 0 \right]}\left( {t,{t_1}} \right){\varphi _{n0}}\left( {{t_1}} \right){X_{00}}\left( {{t_1},{t_i}} \right)}
\end{equation}

\noindent
describes the sub-process containing the following elementary processes: The considered system is initially in the state $\left| {{0_{{t_i}}}} \right\rangle $ remaining in this state with a probability amplitude ${X_{00}}\left( {{t_1},{t_i}} \right)$ until the time $t_1$ when it jumps to the state $\left| {{n_{{t_1}}}} \right\rangle $ with a probability amplitude ${\varphi _{n0}}\left( {{t_1}} \right)$. Then it remains in this state until the final time $t$ with a probability amplitude $X_{nn}^{\left[ 0 \right]}\left( {t,{t_1}} \right)$. Neglecting from the latter the transitions to the state $\left| 0 \right\rangle $ avoids double counting: the transitions $\left( {n \to 0,0 \to n} \right)$ give the same contribution to the total amplitude with the transitions $\left( {0 \to n,n \to 0} \right)$ that have already been taken into account in the amplitude ${X_{00}}$. Leaving the technical details for the Appendix, we add here some important remarks before closing this section:

\begin{itemize}
\item The building blocks of the rearranged GPPA are the elementary transition amplitudes (which we will refer to as ''flips'' onwards)      

\begin{equation}
\label{1.1.13}
{{\Phi _{nm}} = \left\langle {{n_t}} \right|i\hbar {\partial _t}\left| {{m_t}} \right\rangle  = \frac{{\left\langle {{n_t}} \right|i\hbar {\partial _t}{{\hat H}_I}\left( t \right)\left| {{m_t}} \right\rangle }}{{{E_m}\left( t \right) - {E_n}\left( t \right)}}{\rm{   }}~~~~~~\left( {n \ne m} \right)}
\end{equation}

\noindent
on which the standard adiabatic perturbation theory (APT) is also based. Thus, most of the usual constraints for the applicability of APT, such as the absence of level crossings and degeneracy of the bound states, hold also for the reformulated version of GPPA. Furthermore the proposed perturbative expansion should provide accurate results for small elementary transition amplitudes.

\item We must stress on the fact that the indices in the diagonal factors $X_{nn}^{\left[ {{n_1},...} \right]}$ appearing in Eq.~(\ref{1.1.11}) are always discrete and they refer to bound states. The reason is that a state belonging to the continuum is of measure zero in the associated spectrum. However, in the intermediate steps involving sums contributing to this amplitude, the continuous  part (if it exists) necessarily participates through the corresponding integrals.

\item The solution (\ref{1.1.10}) determines ${X_{n0}}\left( t \right)$, and consequently the wave function $\left| {{\psi _t}} \right\rangle $, in terms of the elementary transition amplitudes ${\Phi _{nm}}$ and the diagonal factors $X_{nn}^{\left[ {{n_1},...} \right]}$. It is straightforward to recover the APT result by expanding $X_{nn}^{\left[ {{n_1},...} \right]}$ in powers of the elementary transition amplitudes, the flips. As we shall see in the next section, the loop contributions forming the diagonal amplitudes can be resummed yielding non-trivial exponential factors. 

\item
 As it is known \cite{Landau1932}, the transitions between eigenstates is a genuine non-perturbative phenomenon with the Landau-Zener problem the most classical example for this. The diagonal factors $X_{nn}^{\left[ {n_1,...} \right]}$ account for these non-perturbative contributions as they incorporate the loop contributions coming from all the flips that begin from and terminate at the same  state. It is obvious that if the system's Hilbert space is  finite dimensional, the series (\ref{1.1.10}) terminates. For example, in the Landau-Zener two-state system only the first term is different from zero. However, even for an infinite dimensional system the series (\ref{1.1.10}) converges if $\vert {\Phi _{nm}} \vert \to 0$ for $\vert n - m \vert \to 0$. The reason is that in the truncated sum in Eq.~(\ref{1.1.11}) the quantum numbers ought to be strictly different. 

\end{itemize}

\section{Life Times in GPPA}
\label{life}

The analysis of the previous section pointed out the important role of the diagonal factors in the improved perturbative expansion in Eq.~(\ref{1.1.10}). In the current section we present a technique for resuming the loop contributions that constitute these factors. Being  interested for the role of quantum resonances we explore the occurrence of a finite life time for a time-dependent eigenstate (say $\left| {{n_t}} \right\rangle $) within the framework of GPPA. Expressed in terms of transition amplitudes between states of the temporary basis, the finite life times are related with the probability the considered system, being initially ($t=t_i \to -\infty$) at a certain eigenstate (not necessarily a bound one), to end up at time $t=t_f \to +\infty$ eventually in the same state.  To this end we calculate the quantity

\begin{equation}
\label{1.2.1}
{f_n}\left( t \right) = \langle {n_t}|\hat T{e^{ - \frac{i}{\hbar }\int\limits_{{t_i}}^t {dt\hat H\left( t \right)} }}\left| {{n_{{t_i}}}} \right\rangle  = \langle {n_t}|\left. {{\psi _t}} \right\rangle {\text{ }}.
\end{equation}

\noindent
Using the expansion (\ref{1.1.2}) and Eq.~(\ref{1.1.5}) this amplitude is obtained as the coefficient of the first term in the superposition (\ref{1.1.9}) \cite{Deguchi2005}:

\begin{equation}
\label{1.2.2}
{{f_n}\left( t \right) = {e^{ - \frac{i}{\hbar }\int\limits_{{t_i}}^t {dt'{{\bar E}_n}\left( {t'} \right)} }}\left\langle n \right|\hat T{e^{\frac{i}{\hbar }\int\limits_{{t_i}}^t {dt'\hat \varphi \left( {t'} \right)} }}\left| n \right\rangle  = {e^{ - \frac{i}{\hbar }\int\limits_{{t_i}}^t {dt'{{\bar E}_n}\left( {t'} \right)} }}{X_{nn}}\left( {t,{t_i}} \right)}.
\end{equation}

\noindent
After expanding ${X_{nn}}$ in powers of $\varphi$, it is convenient to introduce the amplitude: 

\begin{equation}
\label{1.2.3}
{{\tilde \Phi _{nm}} = {e^{ - \frac{i}{\hbar }t{\varepsilon _n}}}{\varphi _{nm}}{e^{ + \frac{i}{\hbar }t{\varepsilon _m}}}~~;~~{\varepsilon _n} = \frac{1}{{{t_f} - {t_i}}}\int\limits_{{t_i}}^{{t_f}} {dt} {\bar E_n}\left( t \right) \equiv \left\langle {{{\bar E}_n}\left( t \right)} \right\rangle}
\end{equation}

\noindent
and express the result in terms of the Fourier decomposed functions:

\begin{equation}
\label{1.2.4}
\begin{split}
& {B_{nm}}\left( \nu  \right) = \frac{1}{{\sqrt {2T} }}\int\limits_{ - T}^T {dt} {\tilde \Phi _{nm}}\left( t \right){e^{i\omega \nu t}}{\rm{   }}{\rm{,    }} \\ & {\tilde \Phi _{nm}}\left( t \right) = \frac{1}{{\sqrt {2T} }}\sum\limits_{\nu  =  - \infty }^\infty  {{B_{nm}}} \left( \nu  \right){e^{ - i\omega \nu t}}~~;~~\omega  = \pi /T
\end{split}
\end{equation}

\noindent
To write this decomposition we supposed that the time we observe the system lies in the interval $\left[ {{t_{final}} = T,{t_{initial}} =  - T} \right]$ and the functions relevant to its description behave well enough to admit a Fourier series representation. Of course to get the link to the life time we will finally consider the limit $T \to \infty $. In this limit, for a non-periodic driving, the expressions in (\ref{1.2.4}) turn out to be the corresponding Fourier transforms. If the driving is periodic with period $2\pi /{\omega _0}$ we shall assume that $2T = N2\pi /{\omega _0}$ with $N \to \infty $. In this case 
Eq.~(\ref{1.2.4}) represent the conventional Fourier series. 

Taking these changes into account it is straightforward to find:

\begin{equation}
\label{1.2.5}
\begin{split}
{X_{nn}} = & \sum\limits_{r = 1}^\infty  {\frac{{{{\left( { - 1} \right)}^r}}}{{{{\left( {2T} \right)}^{r/2}}}}} \sum\limits_{{n_1},...,{n_{r - 1}}} {\sum\limits_{{\nu _1},...,{\nu _r} =  - \infty }^\infty } \\ & {{ {\frac{{{B_{n{n_{r - 1}}}}\left( {{\nu _r}} \right)...{B_{{n_1}n}}\left( {{\nu _1}} \right){e^{ - \frac{i}{\hbar }t\left( {\hbar \omega \left( {{\nu _1} + ... + {\nu _r}} \right) + i0} \right)}}}}{{\left( {\hbar \omega \left( {{\nu _1} + ... + {\nu _r}} \right) + i0} \right)...\left( {\hbar \omega {\nu _1} - \left( {{\varepsilon _{{n_1}}} - {\varepsilon _n}} \right) + i0} \right)}}} }}
\end{split}
\end{equation}

The index $r$ appearing in the above equations denotes the expansion order of the exponential in 
Eq.~(\ref{1.2.2}). The general term in this expansion contains an ordered product $\left\langle 0 \right|\hat \varphi \left( {{t_r}} \right)...\hat \varphi \left( {{t_1}} \right)\left| 0 \right\rangle $ in which  we inserted $r - 1$ complete basis sets, and we used the Fourier decomposed version of $\tilde \Phi $. Note that we have also inserted an infinitesimal imaginary term to secure causal propagation and convergence in the limit of ${t_i} =  - T \to  - \infty $. In the truncated amplitude $X_{nn}^{\left[ {{n_1},...} \right]}$ the forbidden intermediate states in the corresponding summations are omitted as explained when this quantity was introduced (see the discussion below Eq.~(\ref{1.1.9})).

\par
Concerning the singularities contained in Eq.~(\ref{1.2.5}) it is evident that a simple pole appears whenever ${\nu _1} + ... + {\nu _r} = 0$. However higher order singularities may also occur. To illustrate this, suppose that one (or more) of the indices ${n_i}$ is $n$. Then a pole exist whenever ${\nu _1} + ... + {\nu _{i - 1}} = 0$, and in combination with ${\nu _1} + ... + {\nu _r} = 0$ a double (or higher) pole may also appear.  To handle this singular behaviour we use the standard technique \cite{Sakurai2014}, isolating those of the terms in Eq.~(\ref{1.2.5}) that give rise to singular behavior and calculating the logarithmic derivative $ - i\hbar {\partial _t}\ln {X_{nn}}\left( t \right)$ which turns out to be finite. Finally, we integrate up to the final time to get:

\begin{equation}
\label{1.2.6}
{{{X_{nn}}\left( {{t_f},{t_i}} \right) = {e^{\frac{i}{\hbar }\left( {{t_f} - {t_i}} \right){\gamma _n}}}}}
\end{equation}

\noindent
with

\begin{equation}
\label{1.2.7}
\begin{split}
& {\gamma _n} = \mathop \sum \limits_{r = 2}^\infty  \gamma _n^{(r)} \\
& \gamma _n^{(r)} = \frac{{{{( - 1)}^r}}}{{{{(2T)}^{r/2}}}}\sum\limits_{{n_1} \ne n,..,{n_{r - 1}} \ne n} {\mathop \sum \limits_{{\nu _1},..,{\nu _r} =  - \infty }^\infty  } {\delta _{\sum\limits_{\ell  = 1}^r {{\nu _\ell }} ,0}} \times \\ & \qquad \times \frac{{{B_{n{n_1}}}({\nu _1}){B_{{n_1}{n_2}}}({\nu _2})...{B_{{n_{r - 1}}n}}({\nu _r})}}{{\left( {{\varepsilon _{{n_{r - 1}}}} - {\varepsilon _n} - \hbar \omega {\nu _r} - i0} \right)...\left( {{\varepsilon _{{n_1}}} - {\varepsilon _n} - \hbar \omega ({\nu _2} + ... + {\nu _r}) - i0} \right)}}
\end{split}
\end{equation}

Expressed in a crude manner, the factor ${\gamma}$ can acquire a positive imaginary part whenever at least one of the sums, appearing in its defining equation, becomes an integral. This would occur if one of the following scenarios takes place: (i) the index $n$ in Eq.~(\ref{1.2.7}) has a continuous part, (ii) the discrete spectrum becomes almost generate, i.e. $\epsilon_n \approx \epsilon_{n'}$ with $n \neq n'$ and (iii) the index $\nu_i$ becomes continuous (dense set of frequencies).
In these cases

\begin{equation}
\label{1.2.8}
{{\left| {{f_n}\left( {{t_f}} \right)} \right|^2} = {\left| {{X_{nn}}\left( {{t_f},{t_i}} \right)} \right|^2} = {e^{ - \frac{2}{\hbar }\left( {{t_f} - {t_i}} \right)\left| {{\mathop{\rm Im}\nolimits} {\gamma _n}} \right|}} \equiv {e^{ - \left( {{t_f} - {t_i}} \right)/{\tau _n}}}}
\end{equation}

\noindent
where ${\tau _n} = \hbar /2{\mathop{\rm Im}\nolimits} {\gamma _n}$ is the life-time of the temporal state $\left| {{n_t}} \right\rangle $. When dealing with bound states, the occurrence of a finite ${\tau _n}{\rm{ }}{\rm{, }}~~\forall n$ is phenomenologically related to the situation when the system being initially in a definite Hamiltonian eigenstate eventually ends up at a superposition of eigenstates. 

\par
The same kind of calculation can be carried out for the truncated amplitude $X_{nn}^{\left[ {{n_1},...} \right]}\left( {{t_b},{t_a}} \right)$ (being defined through an appropriately truncated sum over the basis $\left| n \right\rangle $) and the result has the following form:

\begin{equation}
\label{1.2.9}
{X_{nn}^{\left[ {{n_1},...} \right]}\left( {{t_b},{t_a}} \right) \sim {e^{\frac{i}{\hbar }\left( {{t_b} - {t_a}} \right)\gamma _n^{\left[ {{n_1},...} \right]} + {\rm{oscillating~terms}}}}}
\end{equation}

\noindent
The factor $\gamma _n^{\left[ {{n_1},...} \right]}$ is similar to ${\gamma _n}$ in Eq.~(\ref{1.2.7}) with the difference that the summations over the indices $n_i$ are here appropriately  truncated. For the general  discussion concerning the appearance of finite life times in driven systems within the framework of GPPA the exact form of each one of the factors in Eq.~(\ref{1.2.9}) is not important. The key property is the existence of a positive imaginary part in the factor $\gamma _n^{\left[ {{n_1},...} \right]}$ resulting to the appearance of a non trivial damping factor to the corresponding probability amplitude. 

\par 
To allow for a more transparent presentation of the developed formalism let us write:

\begin{equation}
\label{1.2.10}
X_{nn}^{\left[ {{n_1},...} \right]}\left( {{t_b},{t_a}} \right) \equiv {e^{\frac{i}{\hbar }\int\limits_{{t_a}}^{{t_b}} {d\tau \delta _n^{\left[ {{n_1},...} \right]}} }}
\end{equation}

\noindent
and reformulate the terms in the series expansion (\ref{1.1.10}) as:

\begin{equation}
\label{1.2.11}
\begin{split}
& {{\varphi _{{n_{j + 1}}{n_j}}}\left( {{t_{j + 1}}} \right)X_{{n_j}{n_j}}^{\left[ {{n_{j - 1}},...,0} \right]}{\varphi _{{n_j}{n_{j - 1}}}}\left( {{t_j}} \right) \sim } \\ & { {\Phi _{{n_{j + 1}}{n_j}}}\left( {{t_{j + 1}}} \right) {e^{\frac{i}{\hbar }\int\limits_{{t_j}}^{{t_{j + 1}}}{d\tau \left( {{{\bar E}_{{n_j}}} - \delta _{{n_j}}^{\left[ {{n_{j - 1}},...,0} \right]}} \right)} }} {\Phi _{{n_j}{n_{j - 1}}}}\left( {{t_j}} \right)}
\end{split}
\end{equation}

\noindent
This leads us to a form allowing for a simple interpretation: The system, at the time $t = {t_j}$, jumps from the state $\left| {{n_{j - 1}}} \right\rangle $ to a different state $\left| {{n_j}} \right\rangle $. Being in this state it propagates till the time $t = {t_{j + 1}}$ via the non-trivial ("dressed") propagator  

\begin{equation}
\label{1.2.12}
{\Delta _{{n_j}}^{\left[ {{n_{j - 1}},...,0} \right]R}\left( {{t_{j + 1}},{t_j}} \right) = \frac{i}{\hbar }\theta \left( {{t_{j + 1}} - {t_j}} \right){e^{\frac{i}{\hbar }\int\limits_{{t_j}}^{{t_{j + 1}}} {d\tau \left( {{{\bar E}_{{n_j}}} - \delta _{{n_j}}^{\left[ {{n_{j - 1}},...,0} \right]}} \right)} }}}
\end{equation}

\noindent
and then it jumps to the state $\left| {{n_{j + 1}}} \right\rangle $. The propagator (\ref{1.2.12}), which is called the geometric phase propagator in \cite{Diakonos2012}, has been produced by the inclusion of all the "loop" transitions that begin from the state $\left| {{n_j}} \right\rangle $ and end up at the same state in the considered time interval.  After this analysis, the solution (\ref{1.1.7}) of the original general problem can be cast in the form:

\begin{equation}
\label{1.2.13}
\begin{split}
\left| {{\psi _t}} \right\rangle  = & {X_{00}}\left( t \right)\left| {{0_t}} \right\rangle  - i\hbar \sum\limits_{n \ne 0} {\sum\limits_{r = 1}^\infty  {\sum\limits_{{n_1} \ne ... \ne {n_{r - 1}} \ne 0,n} {\int\limits_{ - \infty }^\infty  {d{t_r}...} } } } \int\limits_{ - \infty }^\infty  {d{t_1}} \times \\ & \times \Delta _n^{[{n_{r - 1}},...0]R}\left( {t,{t_r}} \right){\Phi _{n{n_{r - 1}}}}\left( {{t_r}} \right)\Delta _{{n_{r - 1}}}^{[{n_{r - 2}},...0]R}\left( {{t_r},{t_{r - 1}}} \right) \times \\ & \times ...{\Phi _{{n_1}0}}\left( {{t_1}} \right)\Delta _0^R\left( {{t_1}, - \infty } \right)\left| {{n_t}} \right\rangle
\end{split}
\end{equation}

Before proceeding with specific examples it is worth to note the following: In the limit $T \to \infty $ the (inverse) "time life" $\gamma $, being proportional to $1/T$, seems to be negligible and consequently irrelevant for the analysis of the driven system we are interested for.  However this is not always the case. For example, consider periodic driving of period ${T_0} = 2\pi /{\omega _0}$ and $ - {t_i} = {t_f} = N{T_0},{\rm{ }}N \to \infty $. This periodicity results to the replacement $T \to {T_0}$ and $\omega  \to {\omega _0}$ in the Fourier decomposition expressions (\ref{1.2.4}) making the factor $\gamma $ a finite quantity. When this periodic system is embedded into the continuum, the factor $\gamma$ may acquire a positive imaginary part yielding non trivial resonant behaviour, as we shall see in a concrete example in the next section. However, even in the case of a non periodic driving, $\gamma$ may have a non trivial impact on the asymptotic behavior of a time dependent quantity. This is clear from the general expression (\ref{1.1.11}). If $ - {t_i} = t = T \to \infty $ a non trivial damping factor may appear:

\begin{equation}
\label{1.2.14}
{\left| {\exp \left( {i2T\gamma /\hbar } \right)} \right| \sim \exp \left( { - 2T\left| {Im\gamma } \right|/\hbar } \right)}
\end{equation}

\section{Calculating life times in specific examples}
\label{ex}
\subsection{Driven Harmonic Oscillator}
\label{dho}

Our first example refers to the simplest bound system with explicit time dependence, a driven harmonic oscillator:

\begin{equation}
\label{1.3.1}
{\hat H\left( t \right) = \frac{{{{\hat p}^2}}}{{2m}} + \frac{1}{2}m{\Omega ^2}{\hat x^2} + \hat xJ\left( t \right)}
\end{equation}

\noindent
Being exactly solvable \cite{Gilbey1966}, it is ideally suited for demonstrating the "improved" perturbative approach developed in the previous section. Using the Hamiltonian (\ref{1.3.1}), it is very easy to find the relevant quantities we are interested for:

\begin{equation}
\label{1.3.2}
{{E_n}\left( t \right) = \hbar \Omega \left( {n + \frac{1}{2}} \right) - \frac{{{J^2}\left( t \right)}}{{2m{\Omega ^2}}}~~;~~{\varepsilon _n} = \hbar \Omega \left( {n + \frac{1}{2}} \right) - \frac{1}{{2m{\Omega ^2}}}\left\langle {{J^2}\left( t \right)} \right\rangle}
\end{equation}

\begin{equation}
\label{1.3.3}
{{\Phi _{nm}}\left( t \right) = \frac{i}{\Omega }{\left( {\frac{\hbar }{{2m\Omega }}} \right)^{1/2}}\left( {\sqrt {n + 1} {\delta _{n,m - 1}} - \sqrt n {\delta _{n,m + 1}}} \right)\dot J\left( t \right) \equiv i{A_{nm}}\dot J\left( t \right)}
\end{equation}

Since the temporary energy spectrum of the oscillator is always discrete it is not possible to get finite life times unless we introduce, through the driving term, some continuous parameters leading to the appearance of integrals in the expression (\ref{1.2.7}). Then, according to the discussion in the previous section, a finite imaginary part in the corresponding $\gamma$ may emerge when the driving is polychromatic involving a dense spectrum of random frequencies. Here we consider the driving to be a superposition of a large number of pulses of the form: 

\begin{equation}
\label{1.3.4}
J\left( t \right) = g\frac{1}{N}\sum\limits_{j = 1}^N {\sin \left( {{\omega _j}t} \right)} , ~~~N \to \infty
\end{equation}

\noindent
Furthermore we assume that the random frequencies ${\omega _j}$ are gaussian distributed random variables centered around $\Omega$ following the probability density $p\left( {{\omega}}\right)$ given by: 

\begin{equation}
\label{1.3.5}
{p\left( {{\omega}} \right) \sim {e^{ - {{\left( {{\omega} - \Omega } \right)}^2}/2{\sigma ^2}}},~~~\sigma  \gg \Omega}
\end{equation}

\noindent
where $\sigma$ is the corresponding variance. Note that the requirement $\sigma  \gg \Omega$ in Eq.~(\ref{1.3.5}) ensures the validity of the perturbative expansion as it will be shown in the following. However, especially for the harmonic potential case, the perturbation series can be easily resummed leading to an exact analytic result without invoking this constrain. Of course our primary goal is to develop a scheme applicable to a general bound system, therefore we will focus exclusively on the perturbative treatment. 

\par
The first order contribution to ${\gamma _n}$ as it appears in Eq.~(\ref{1.2.7}) reads:

\begin{equation}
\label{1.3.6}
\begin{split}
2T\gamma _n^{\left( 2 \right)}{\text{ }} & = \sum\limits_{\nu  =  - \infty }^\infty  {\sum\limits_{n_1} {\frac{{{B_{n{n_1}}}\left( \nu  \right){B_{{n_1}n}}\left( { - \nu } \right)}}{{{\varepsilon _{n_1}} - {\varepsilon _n} + \hbar \omega \nu  - i0}}} }= \\ & = \sum\limits_{n_1} {\int\limits_{ - \infty }^\infty  {d{t_2}} } \int\limits_{ - \infty }^{{t_2}} {d{t_1}{\varphi _{nm}}\left( {{t_2}} \right)} {\varphi _{{n_1}n}}\left( {{t_1}} \right),~~~~T \to \infty
\end{split}
\end{equation}

\noindent
A straightforward calculation yields the result:

\begin{equation}
\label{1.3.7}
{2T\gamma _n^{\left( 2 \right)} = i\hbar \left( {n + \frac{1}{2}} \right){a^2},~~~~a = {\left( {\frac{\pi }{{\hbar m\Omega }}{{\left| {\tilde J\left( \Omega  \right)} \right|}^2}} \right)^{1/2}}}
\end{equation}

\noindent
where

\begin{equation}
\label{1.3.8}
{\tilde J\left( k \right) = \frac{1}{{2i}}\frac{g}{\sigma }\left[ {{e^{ - {{\left( {k + \Omega } \right)}^2}/2{\sigma ^2}}} - {e^{ - {{\left( {k - \Omega } \right)}^2}/2{\sigma ^2}}}} \right]}
\end{equation}

\noindent
is the Fourier transform of the driving (\ref{1.3.4}) in the limit $N \to \infty $. For the case at hand the third order term in (\ref{1.2.7}) does not contribute and the next to leading order correction comes from the fourth order term:

\begin{equation}
\label{1.3.9}
{2T\gamma_n^{\left( 4 \right)} = \frac{{i\hbar }}{4}n\left( {n + 1} \right){a^4}}
\end{equation}

\noindent
For the ground state this term vanishes and it is a simple exercise to verify that, due to the form of ${A_{nm}}$, the same holds for all the higher order contributions to $n=0$. Thus, for the ground state the exact result reads:

\begin{equation}
\label{1.3.10}
{2T{\mathop{\rm Im}\nolimits} {\gamma _0} = \frac{\hbar }{2}{a^2}}
\end{equation}

\noindent
Consequently the probability a system that initially is at a certain eigenstate to end up eventually at the same state can be written in the general form

\begin{equation}
\label{1.3.11}
{{\left| {{X_{nn}}\left( {\infty , - \infty } \right)} \right|^2} = f\left( {n,{a^2}} \right){e^{ - {a^2}}}}
\end{equation}

\noindent
where the function $f\left( n \right)$ with $f\left( 0 \right) = 1$ has to be calculated perturbatively. Using Eqs.~(\ref{1.3.7}) and (\ref{1.3.9}) we find:

\begin{equation}
\label{1.3.12}
{f\left( {n,{a^2}} \right) = 1 - 2n{a^2} + \frac{n}{2}\left( {3n - 1} \right){a^4} + O\left( {{a^6}} \right)}
\end{equation}

It is worth noting that by expanding the exponential factor and using  Eq.(\ref{1.3.12}), the probability  in Eq.~(\ref{1.3.11}) coincides with the standard perturbative result \cite{Gilbey1966}. 

\par
As dictated by Eq. (\ref{1.3.3}), in this example where the time dependent coupling is linear, the flips $\Phi _{n0}(t)$ (for their definition see Eq. (\ref{1.1.13})) within our approach are arranged so that they connect only next neighbours. Then  Eq.~(\ref{1.1.10}) reads:

\begin{equation}
\label{1.3.13}
{{X_{n0}}\left( {t,{t_i}} \right) = \sum\limits_{r = n}^\infty  {X{{_{n0}^{\left( r \right)}}^R}\left( {t,{t_i}} \right)}}
\end{equation}

To illustrate how the terms ${X_{n0}}$ are calculated in practice let us discuss as a concrete example the case $n=1$. In Eq.~(\ref{1.1.11}) can be readily seen that the indexing of all the terms besides the first one can be written as:

\begin{equation}
\label{1.3.14}
{X{_{10}^{\left( r \right)R}} \sim \sum\limits_{{n_{r - 1}} \ne ... = {n_1} \ne 0,1} {{A_{1{n_{r - 1}}}}...{A_{{n_1}0}}}}
\end{equation}

\noindent
Employing Eq.~(\ref{1.3.3}) we can directly see that all higher order terms vanish leaving only the first one, which is actually the exact result:

\begin{equation}
\label{1.3.15}
{X_{10}^{\left( 1 \right)R}\left( {t, - T} \right) = \frac{i}{\hbar }\int\limits_{ - T}^t {d{t_1}} X_{11}^{\left[ 0 \right]}\left( {t,{t_1}} \right){\Phi _{10}}\left( {{t_1}} \right){e^{i\Omega {t_1}}}{X_{00}}\left( {{t_1}, - T} \right){~~~~~}\left( {T \to \infty } \right)}
\end{equation}

\noindent
The general form of the factors $X$ can be read in Eq.~(\ref{1.2.9}) where the oscillating terms get strongly suppressed in the limit $T \to \infty $:  

\begin{equation}
\label{1.3.16}
{{X_{00}}\left( {{t_1}, - T} \right) = {e^{\frac{i}{\hbar }\left( {{t_1} + T} \right){\gamma _0} + O\left( {1/T} \right)}}\mathop  = \limits_{T \to \infty } {e^{\frac{i}{\hbar }\left( {{t_1} + T} \right){\gamma _0}}}}
\end{equation}

\noindent
To obtain the last result, we took into account that $\gamma  \sim 1/T$. In addition, to obtain the asymptotic behaviour of the amplitude (\ref{1.3.15}) we assumed that the upper limit in the integration over the time ${t_1}$ approaches the value $T$. In a similar way we find that

\begin{equation}
\label{1.3.17}
{X_{11}^{\left[ 0 \right]}\left( {t,{t_1}} \right) = {e^{\frac{i}{\hbar }\left( {t - {t_1}} \right)\gamma _1^{\left[ 0 \right]} + O\left( {1/T} \right)}}\mathop  = \limits_{T \to \infty } {e^{\frac{i}{\hbar }\left( {t - {t_1}} \right)\gamma _1^{\left[ 0 \right]}}}}
\end{equation}

\noindent
Inserting Eqs.~(\ref{1.3.16}) and (\ref{1.3.17}) into Eq.~(\ref{1.3.15}) we get:

\begin{equation}
\label{1.3.18}
{X_{10}^{\left( 1 \right)R}\left( {T, - T} \right) = \frac{i}{\hbar }{e^{\frac{i}{\hbar }T\left( {{\gamma _0} + \gamma _1^{\left[ 0 \right]}} \right)}}\int\limits_{ - T}^T {d{t_1}} {\Phi _{10}}\left( {{t_1}} \right){e^{i\Omega {t_1} + \frac{i}{\hbar }{t_1}\left( {{\gamma _0} - \gamma _1^{\left[ 0 \right]}} \right)}}}
\end{equation}

\noindent
The time integration is performed by invoking the Fourier transform of the amplitude $\Phi $:

\begin{equation}
\label{1.3.19}
{{B_{10}}\left( k \right) = \int\limits_{ - \infty }^\infty  {\frac{{dt}}{{\sqrt {2\pi } }}} {\Phi _{10}}\left( t \right){e^{ikt}} =  - \frac{1}{\Omega }{\left( {\frac{\hbar }{{2m\Omega }}} \right)^{1/2}}k\tilde J\left( k \right)}
\end{equation}

\noindent
Using the last expression we find:

\begin{equation}
\label{1.3.20}
{X_{10}^{\left( 1 \right)R}\left( {T, - T} \right) =  - a{e^{\frac{i}{\hbar }T\left( {{\gamma _0} + \gamma _1^{\left[ 0 \right]}} \right)}}{e^{\frac{i}{\hbar }T\left( {{\gamma _0} - \gamma _1^{\left[ 0 \right]}} \right)}}{\rm{ = }} - a{e^{\frac{i}{\hbar }2T{\gamma _0}}}}
\end{equation}

\noindent
Finally, employing (\ref{1.3.10}) yields the following exact result for the amplitude ${X_{10}}$:

\begin{equation}
\label{1.3.21}
{X_{10}^{\left( 1 \right)R}\left( {T, - T} \right) =  - a{e^{\frac{i}{\hbar }T\left( {{\gamma _0} + \gamma _1^{\left[ 0 \right]}} \right)}}{e^{\frac{i}{\hbar }T\left( {{\gamma _0} - \gamma _1^{\left[ 0 \right]}} \right)}}{\rm{ = }} - a{e^{\frac{i}{\hbar }2T{\gamma _0}}}}
\end{equation}

\noindent
Each of the renormalized coefficients in Eq.~(\ref{1.1.10}) is treated in an analogous manner leading to a similar damping factor as the one appearing in Eq.~(\ref{1.3.21}). Clearly this factor is inherited to the total amplitude:

\begin{equation}
\label{1.3.22}
{{X_{n0}}\left( {\infty , - \infty } \right) \sim {e^{ - {a^2}/2}}}
\end{equation}

As already discussed the result of Eq.~(\ref{1.3.21}) is the exact one \cite{Gilbey1966} to be compared with the corresponding APT result which is equal to $- a(1 + 0({a^2}))$.

\begin{figure}[ht]
    \centering
        \includegraphics[width=1.00\textwidth]{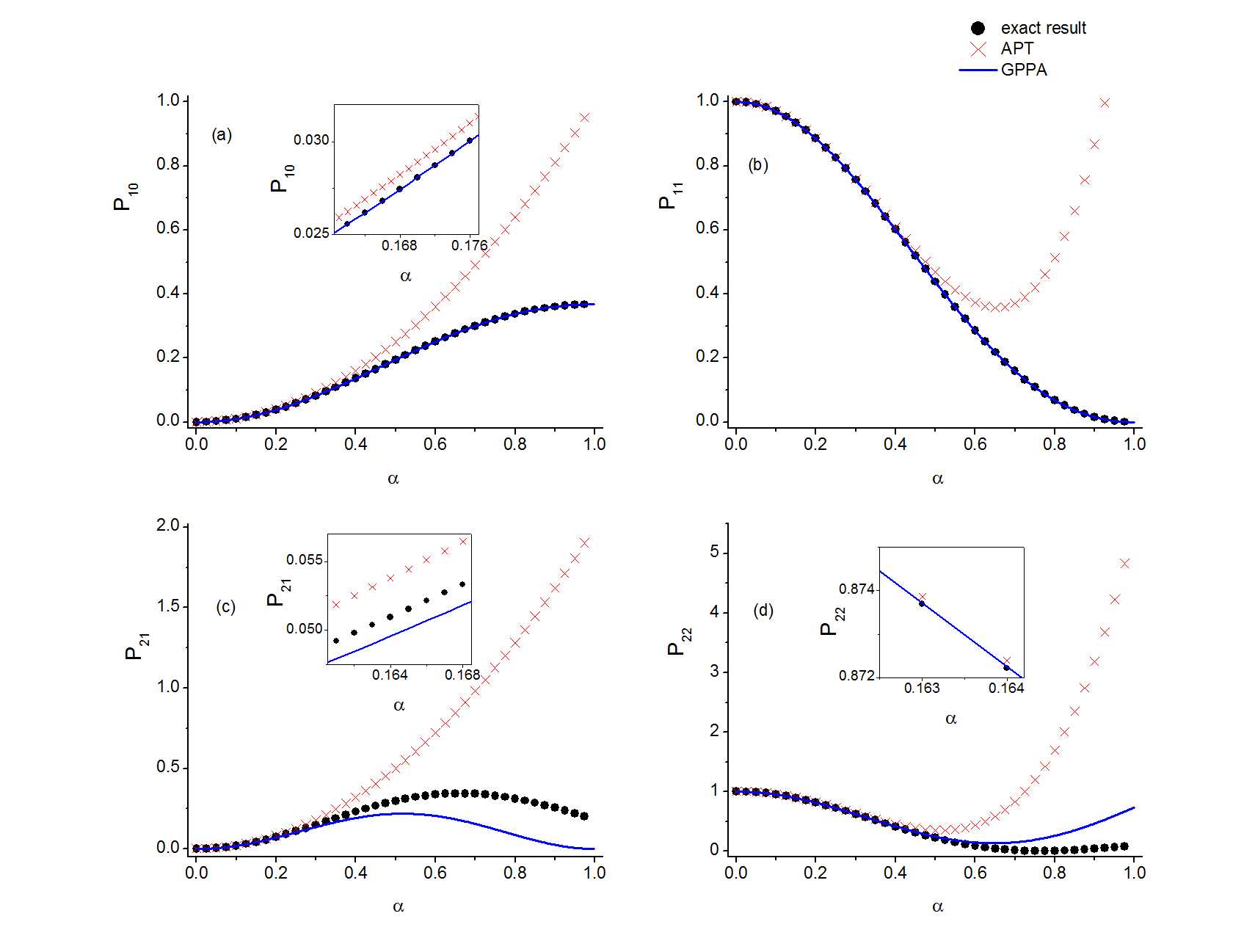}
        \vspace{-1.3cm}
    \caption{Various transition probabilities for the DHO as a function of $\alpha$ where GPPA is compared to APT (both calculated up to fourth order) and the exact result. In the cases of ${P_{10}}$ (a) and ${P_{11}}$ (b) (and all ${P_{n0}}$ as mentioned in the main text) GPPA recovers the exact result. Also for ${P_{21}}$ (c) and ${P_{22}}$ (d) our approach gives better results than the traditional APT due to the inclusion of the dumping factor (\ref{1.3.22}). Note that the inset in the (b) case has similar scales as the (d) one and therefore is omitted.}
    \label{fig:dho}
\end{figure}

\par
In order to better demonstrate our method we plot in Figure \ref{fig:dho} various cases of transition probabilities ${P_{nm}} \equiv {\left| {{X_{nm}}\left( {\infty , - \infty } \right)} \right|^2}$ as a function of the parameter $\alpha$ where the exact result is compared to our method and APT (both calculated up to fourth order). In Figure \ref{fig:dho}a we plot the transition probability for flipping to the first excited state starting from the ground one (${P_{10}}$). In this case the exact result is recovered by GPPA and we can see in the relative inset the difference from the APT for relatively small values of $\alpha$ ($\alpha<0.2$) deviating at most by $4\%$. In Figure \ref{fig:dho}b we show the return probability to the first excited state (${P_{11}}$). In this case the exact result is also recovered by GPPA and the difference with APT is at least one order of magnitude smaller in the low $\alpha$ regime than the case of Figure \ref{fig:dho}a (the corresponding inset is omitted). In Figure \ref{fig:dho}c we plot the the transition probability for flipping to the second excited state starting from the first one (${P_{21}}$). We observe that the result obtained with our approach is closer to the exact one than the APT result even for small $\alpha$ values (see corresponding inset). Finally in Figure \ref{fig:dho}d we plot the return probability to the second excited state (${P_{22}}$). It is again clear that our method gives better results. For small $\alpha$ the diferrence between the two method is less than $1$\textperthousand. In the two latter cases our method gives significally better results due to the inclusion of the dumping factor (\ref{1.3.22}).

Note that the renormalized series expansion for the amplitude $X_{nm}$, $n \neq m$   always terminates for the harmonic oscillator case as long as the coupling between the system and the driving is of the form $\sum_{j=1}^N a_j x^j$, $N < \infty$. For the linear driving in Eq.~(\ref{1.3.1})) it is easily checked that the number of terms in the series $X_{nm}$ is  equal to min$(n+1,m+1)$.

\subsection{A periodically driven delta function}
\label{delta}

In the previous example we examined a simple bound system under the influence of a non periodic driving, and we verified that the proposed approach improves the usual perturbation theory. However, when the system in consideration is embedded into the continuum and the driving is periodic, our approach can reveal further non trivial phenomena. The reason is that in this case, as we have already noted, the life time of a bound state becomes finite leading to interesting quantum resonant behaviour which in turn influences transport properties of the considered system.

Before discussing a specific example, it is useful to revisit Eq.~(\ref{1.2.7}) trying to rearrange suitably the terms defining ${\gamma _n}$. Following essentially the same technique as that used to construct the renormalized version of the transition amplitudes we first isolate in the fourth term of Eq.~(\ref{1.2.7}) the contribution coming from ${n_3} = {n_1},{\rm{ }}{\nu _3} + {\nu _2} = 0$ recombining it with the first term to get:

\begin{equation}
\label{1.3.23}
\begin{split}
& \sum\limits_{n_1} {\sum\limits_{{\nu _1},{\nu _2} =  - \infty }^\infty  {\frac{{{B_{nn_1}}\left( {{\nu _1}} \right){B_{n_1n}}\left( { - {\nu _1}} \right)}}{{{\varepsilon _{n_1}} - {\varepsilon _n} + \hbar \omega {\nu _1} - i0}}} } \to  \\ &  {\text{ }}\sum\limits_{n_1} {\sum\limits_{{\nu _1} =  - \infty }^\infty  {\frac{{{B_{nn_1}}\left( {{\nu _1}} \right){B_{n_1n}}\left( { - {\nu _1}} \right)}}{{{\varepsilon _{n_1}} - {\varepsilon _n} + \hbar \omega {\nu _1} - i0}}} } \times \\ & \times \left[ 1  + \frac{1}{{2T}}\frac{1}{{{\varepsilon _{n_1}} - {\varepsilon _n} + \hbar \omega {\nu _1} - i0}}\sum\limits_{{n_2} \ne n} {\sum\limits_{{\nu _2} =  - \infty }^\infty  {\frac{{{B_{n_1{n_2}}}\left( {{\nu _2}} \right){B_{{n_2}n_1}}\left( { - {\nu _2}} \right)}}{{{\varepsilon _{{n_2}}} - {\varepsilon _n} + \hbar \omega \left( {{\nu _1} + {\nu _2}} \right) - i0}}}  } \right]
\end{split}
\end{equation}

\noindent
Then we redefine all the terms appearing in Eq.~(\ref{1.2.7}), including also higher order corrections, in the following manner:

\begin{equation}
\label{1.3.24}
{{\gamma _n} = \frac{1}{{2T}}\sum\limits_{n_1} {\sum\limits_{{\nu _1} =  - \infty }^\infty  {\frac{{{B_{nn_1}}\left( {{\nu _1}} \right){B_{{n_1}n}}\left( { - {\nu _1}} \right)}}{{{\varepsilon _{n_1}} - {\varepsilon _n} - \delta {\varepsilon _{{n_1}n}}\left( {{\nu _1}} \right) + \hbar \omega {\nu _1} - i0}}} }  + O\left( {{B^3}} \right)}
\end{equation}

\noindent
The leading contribution to the "energy correction" term in the denominator of the last expression reads:

\begin{equation}
\label{1.3.25}
{\delta {\varepsilon _{{n_1}n}}\left( {{\nu _1}} \right) = \frac{1}{{2T}}\sum\limits_{{n_2} \ne n} {\sum\limits_{{\nu _2} =  - \infty }^\infty  {\frac{{{B_{n_1{n_2}}}\left( {{\nu _2}} \right){B_{{n_2}n_1}}\left( { - {\nu _2}} \right)}}{{{\varepsilon _{{n_2}}} - {\varepsilon _n} + \hbar \omega \left( {{\nu _1} + {\nu _2}} \right) - i0}}} }  + O\left( {{B^3}} \right)}
\end{equation}

Let us make some clarifying remarks at this point:
\begin{itemize}
\item{Notice that in the example considered in this subsection ${\gamma _n}$ and $\delta {\varepsilon _{{n_1}n}}$ are not negligible as it was the case in the example of the previous subsection where the driving was non periodic.}  

\item{While the index $n$ (see Eq. (\ref{1.3.24})) may belong to the discrete or to the continuum part of the spectrum, the index $n_1$ that defines the energy correction is always discrete. Obviously when the indices ${n_1}, n$ belong to the continuum the constrain ${n_1} \ne n$ is of measure zero and becomes irrelevant.}

\item{We could perform a similar resummation for the non periodically driven system considered in the previous subsection. However, in that case, the energy correction, being proportional to $1/T$, is negligible for $T \to \infty$. Thus in the first example we focused on the calculation of time dependent quantities in the limit $t \to T$ where a non trivial damping factor could appear. However, when we are dealing with time independent quantities like  ${\gamma _n}$, a resummation like the one in Eq.~(\ref{1.3.24}) has no practical meaning for a non-periodic system.}
\end{itemize}

\par
In the following we will apply the GPPA formalism for life time estimation, as developed in the previous sections, to our second example concerning quantum evolution in a driven delta-barrier described by the Hamiltonian: 

\begin{equation}
\label{1.3.26}
{\hat H = \frac{{{{\hat p}^2}}}{{2m}} - {g_0}\delta \left( x \right)\sin \omega t}
\end{equation}

\noindent
In \cite{Diakonos2012} we used GPPA to interpret the transmission properties in this potential. Here we focus on the emergence of a finite life time for the single bound state contained in the temporary energy spectrum. We avoid to present the details of the solution to the instantaneous problem (which can be found in \cite{Diakonos2012}) and we proceed directly to the calculation of the amplitude:

\begin{equation}
\label{1.3.27}
{{X_{kk}}\left( {T, - T} \right) \sim {e^{i2T{\gamma _k}}},~~~T = \pi N}
\end{equation}

\noindent
for momentum $k$ in the continuous part of the Hamiltonian spectrum and the associated factor ${\gamma _k}$ for a state belonging to the continuum. Such a quantity is naturally related to the transmission amplitude in a scattering process. For a certain incoming state, the ratio:

\begin{equation}
\label{1.3.28}
{{T_{kk}} \equiv {X_{kk}}\left( {\infty , - \infty } \right)/X_{kk}^{\left[ 0 \right]}\left( {\infty , - \infty } \right),~~~\left| {{T_{kk}}} \right| \le 1}
\end{equation}

\noindent
is a measure of the impact of the bound state on the scattering process: to be more explicit, if $\left| {{T_{kk}}} \right| \approx 1$ the existence of the bound state does not significantly influence the corresponding transmission coefficient which is essentially controlled by flips between states belonging to the continuum. As $\left| {{T_{kk}}} \right|$ decreases, the influence of the bound state on the transmission sets in, leading to a total suppression of the transmission when $\left| {{T_{kk}}} \right| \to 0$. This is phenomenologically interpreted as the emergence of a Fano resonance in \cite{Diakonos2012}. In the following we will show how this suppression is related to the life time of the emerging quasi-bound state.

\par
We first consider the leading contribution to ${\gamma _k}$:

\begin{equation}
\label{1.3.29}
\begin{split}
{\gamma _k} = & \frac{1}{{2\pi }}\sum\limits_{n_1} {\sum\limits_{{\nu _1} =  - \infty }^\infty  {\frac{{{B_{k{n_1}}}\left( {{\nu _1}} \right){B_{{n_1}k}}\left( { - {\nu _1}} \right)}}{{{\varepsilon _{n_1}} - {\varepsilon _k} - \delta {\varepsilon _{{n_1}k}}\left( {{\nu _1}} \right) + {\nu _1} - i0}}} }  + O\left( {{B^3}} \right) = \\ & = \frac{1}{{2\pi }}\sum\limits_{{\nu _1} =  - \infty }^\infty  {\frac{{{B_{k0}}\left( {{\nu _1}} \right){B_{0k}}\left( { - {\nu _1}} \right)}}{{{\varepsilon _0} - {\varepsilon _k} - \delta {\varepsilon _{0k}}\left( {{\nu _1}} \right) + {\nu _1} - i0}}} \\ & + \frac{1}{\pi }\sum\limits_{{\nu _1} =  - \infty }^\infty  {\int\limits_0^\infty  {dk'} \frac{{{B_{kk'}}\left( {{\nu _1}} \right){B_{k'k}}\left( { - {\nu _1}} \right)}}{{{\varepsilon _{k'}} - {\varepsilon _k} + {\nu _1} - i0}}}  + O\left( {{B^3}} \right)
\end{split}
\end{equation}

\noindent
The Fourier transformed functions entering in the last equation are:

\begin{equation}
\label{1.3.30}
{{B_{0k}}\left( \nu  \right) = \int\limits_0^\pi  {\frac{{d\tau }}{{\sqrt {2\pi } }}} {\Phi _{0k}}\left( \tau  \right){e^{\frac{i}{8}{g^2}\sin 2\tau }}{e^{i\nu \tau }},{\rm{    }}{B_{k'k}}\left( \nu  \right) = \int\limits_{ - \pi }^\pi  {\frac{{d\tau }}{{\sqrt {2\pi } }}} {\Phi _{k'k}}\left( \tau  \right){e^{i\nu \tau }}}
\end{equation}

\noindent
The leading contribution to the energy correction term in Eq.~(\ref{1.3.29}) is found in Eq.~(\ref{1.3.25}):

\begin{equation}
\label{1.3.31}
{\delta {\varepsilon _{0k}}\left( {{\nu _1}} \right) = \sum\limits_{{\nu _2} =  - \infty }^\infty  {\int\limits_0^\infty  {\frac{{dk'}}{\pi }} } \frac{{{B_{0k'}}\left( {{\nu _2}} \right){B_{k'0}}\left( { - {\nu _2}} \right)}}{{{\varepsilon _{k'}} - {\varepsilon _k} + {\nu _1} + {\nu _2} - i0}} + O\left( {{B^3}} \right)}
\end{equation}

\noindent
Non-trivial behaviour originating from the bound state is expected to occur whenever the initial (:incoming ) energy is

\begin{equation}
\label{1.3.32}
{{\varepsilon _k} = {k^2}/2 \approx {\varepsilon _0} + n}
\end{equation}

\noindent
where $n$ must  be a strictly positive integer since ${\varepsilon _0}$ is negative. In such a case:

\begin{equation}
\label{1.3.33}
\begin{split}
{\gamma _k} = & \frac{1}{{2\pi }}\frac{{{B_{k0}}\left( n \right){B_{0k}}\left( { - n} \right)}}{{ - \delta {\varepsilon _{0k}}\left( n \right)}} + \frac{1}{{2\pi }}\sum\limits_{{\nu _1} \ne n}^\infty  {\frac{{{B_{k0}}\left( {{\nu _1}} \right){B_{0k}}\left( { - {\nu _1}} \right)}}{{{\nu _1} - n - \delta {\varepsilon _{0k}}\left( {{\nu _1}} \right)}}}  \\ & { + \frac{1}{\pi }\sum\limits_{{\nu _1} =  - \infty }^\infty  {\int\limits_0^\infty  {dk'} \frac{{{B_{kk'}}\left( {{\nu _1}} \right){B_{k'k}}\left( { - {\nu _1}} \right)}}{{{\varepsilon _{k'}} - {\varepsilon _k} + {\nu _1} - i0}}}  + O\left( {{B^3}} \right)}
\end{split}
\end{equation}

\noindent
Note that when ${\varepsilon _k} \ne {\varepsilon _0} + {\rm{integer}}$, it is not correct to take into account the energy correction contribution in the denominator of the first term in the rhs of Eq.~(\ref{1.3.29}), since other terms of the same order ($O\left( {{B^3}} \right)$) have been omitted in our expansion. For similar reasons, the energy correction in the second term in the rhs of the last equation, must also be neglected, making this term real. Therefore it does not contribute in the leading order calculation of $\left| {{T_{kk}}} \right|$. The third term, being part of the amplitude $X_{kk}^{\left[ 0 \right]}$, will be cancelled out in the final result. However, within the framework of a perturbative calculation it may happen that this term becomes of lower order than the first one. In this case, if dictated by the order of the perturbative calculation, one has to keep this term and omit the first one. 

\par
Applying Eq.~(\ref{1.3.31}) for ${\nu _1} = n$ we find:

\begin{equation}
\label{1.3.34}
\begin{split}
\delta {\varepsilon _{0k}}\left( n \right) & = \sum\limits_{{\nu _2} =  - \infty }^\infty  {\int\limits_0^\infty  {\frac{{dk'}}{\pi }} } \frac{{{B_{0k'}}\left( {{\nu _2}} \right){B_{k'0}}\left( { - {\nu _2}} \right)}}{{{\varepsilon _{k'}} - {\varepsilon _k} + n + {\nu _2} - i0}} + O\left( {{B^3}} \right) = \\ & = \sum\limits_{{\nu _2} =  - \infty }^\infty  {\int\limits_0^\infty  {\frac{{dk'}}{\pi }} } \frac{{{B_{0k'}}\left( {{\nu _2}} \right){B_{k'0}}\left( { - {\nu _2}} \right)}}{{{\varepsilon _{k'}} - {\varepsilon _0} + {\nu _2} - i0}} + O\left( {{B^3}} \right)
\end{split}
\end{equation}

\noindent
When the incoming energy has the value (\ref{1.3.32}), the energy correction coincides with the function ${\gamma _0}$ that determines the time life of the bound state: ${\gamma _0} \sim 1/{\tau _0}$. In Eq.~(\ref{1.3.29}) we see that the energy ${\varepsilon _0}$ (the mean value of the time-dependent bound state energy) is "corrected" due to bound-continuum-bound flips. This correction $\delta {\varepsilon _{0k}}$ depends on the incoming energy and in general does not coincide with the energy width ${\gamma _0}$ acquired by the ground state due to its embedding into the continuum. As displayed by the second term on the rhs of Eq.~(\ref{1.3.33}) $\delta {\varepsilon _{0k}}$ becomes important whenever $\nu_1=n$ in the denominator, contributing significantly to 
$\gamma_k$ and therefore to the transmission amplitude. Furthermore, when $\delta {\varepsilon _{0k}} = {\gamma _0}$ then the first term in Eq.~(\ref{1.3.33}) acquires a large positive imaginary part yielding resonant behaviour. In physical terms this emerges when the time needed for the incoming particle for passing through the driven delta barrier, determined by the "dressed" energy of the bound state, coincides with the associated life time.

\par
The correction (\ref{1.3.34}) has a real part:

\begin{equation}
\label{1.3.35}
{{\mathop{\rm Re}\nolimits} \delta {\varepsilon _{0k}}\left( n \right) = \sum\limits_{\nu  =  - \infty }^\infty  {\Pr .\int\limits_0^\infty  {\frac{{dk'}}{\pi }} } \frac{{{{\left| {{B_{0k'}}\left( \nu  \right)} \right|}^2}}}{{{\varepsilon _{k'}} - {\varepsilon _k} + n + \nu }} + O\left( {{B^3}} \right)}
\end{equation}

\noindent
and a positive imaginary part:

\begin{equation}
\label{1.3.36}
{{\mathop{\rm Im}\nolimits} \delta {\varepsilon _{0k}}\left( n \right) = \sum\limits_{\nu  =  - \infty }^\infty  {\int\limits_0^\infty  {dk'} } {\left| {{B_{0k'}}\left( \nu  \right)} \right|^2}\delta \left( {{\varepsilon _{k'}} - {\varepsilon _k} + \left( {n + \nu } \right)} \right) + O\left( {{B^3}} \right)}
\end{equation}

\noindent
Using Eq. (\ref{1.3.30}) we can verify that ${B_{0k}}\left( \nu  \right)$ decreases rapidly with increasing $k$ and $\left| \nu  \right|$. The leading contribution to the sum (\ref{1.3.36}) is obtained for $n=1$ and $\nu  =  - 1$:

\begin{equation}
\label{1.3.37}
{{\mathop{\rm Im}\nolimits} \delta {\varepsilon _{0k}}\left( 1 \right) \approx {\left| {{B_{k0}}\left( 1 \right)} \right|^2}/k}
\end{equation}

\noindent
If the incoming energy is 

\begin{equation}
\label{1.3.38}
{{\varepsilon _k} = {\varepsilon _0} + 1 + {\mathop{\rm Re}\nolimits} \delta {\varepsilon _{0k}}\left( 1 \right)}
\end{equation}

\noindent
the denominator in the first term in the rhs of Eq. (\ref{1.3.33}) becomes purely imaginary:

\begin{equation}
\label{1.3.39}
{\gamma _k}={\frac{i}{{2\pi }}\frac{{{{\left| {{B_{k0}}\left( 1 \right)} \right|}^2}}}{{{\mathop{\rm Im}\nolimits} \delta {\varepsilon _{0k}}\left( 1 \right)}} = i\frac{k}{{2\pi }}}
\end{equation}

\begin{figure}[ht]
    \centering
        \includegraphics[width=1.00\textwidth]{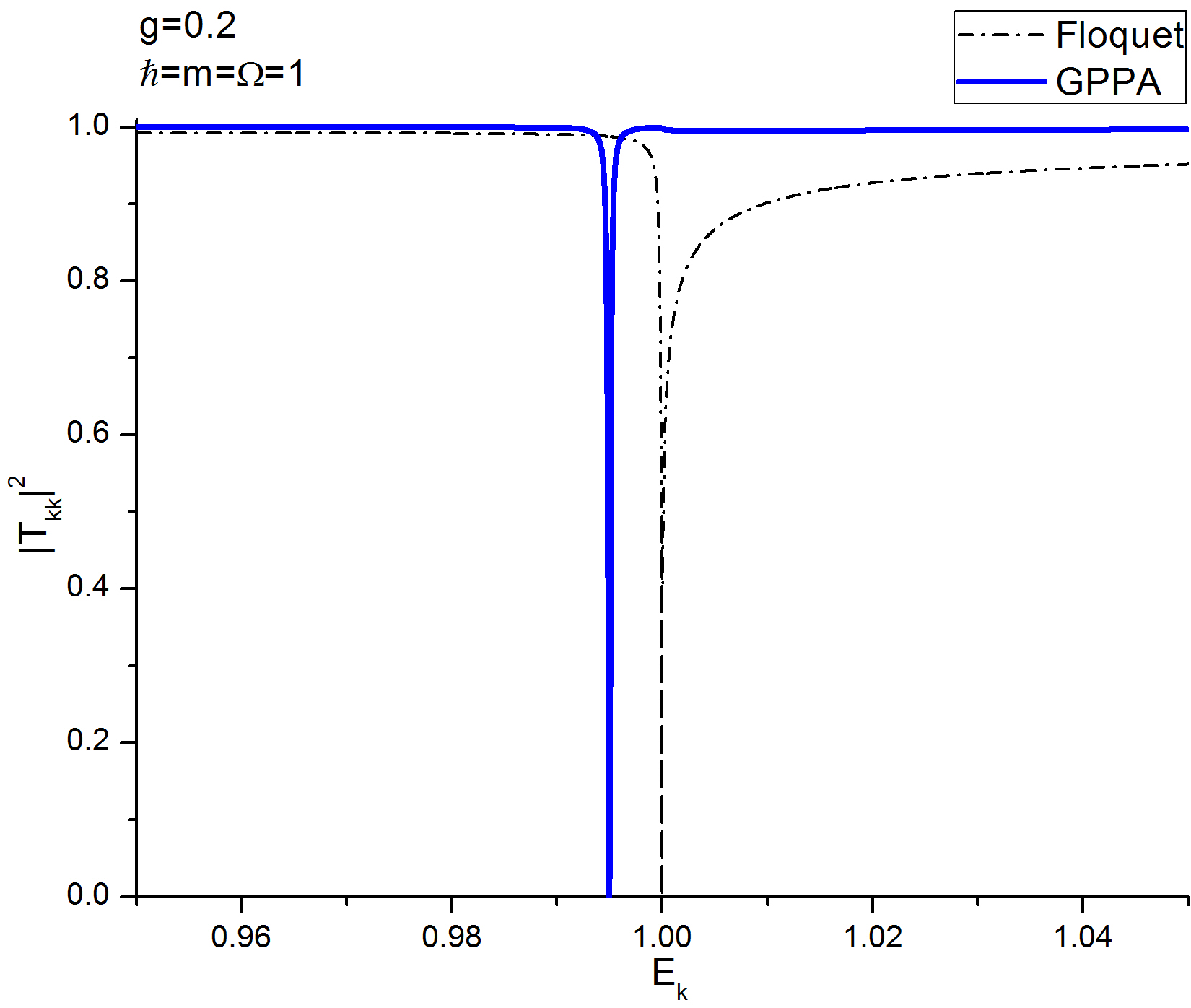}
        \vspace{-0.9cm}
    \caption{Elastic part of the transmission coefficient as a function of incoming energy in the region around the transmission zero.}
    \label{fig:floquet}
\end{figure}

\noindent
Thus, for this specific value of the incoming energy, a zero of the transmission amplitude occurs:

\begin{equation}
\label{1.3.40}
{\left| {{T_{kk}}} \right| \sim {e^{-Nk}}\mathop  \to \limits_{N \to \infty } 0}
\end{equation}

\noindent
This is the condition for the emergence of a Fano resonance in the associated transmission profile as discussed in \cite{Diakonos2012}. The Fano resonance can be clearly seen in Figure \ref{fig:floquet} where the exact result, calculated numerically with Floquet theory \cite{Reichl2001} is compared with the result of GPPA calculated as described above. Note that the APT result cannot reproduce this behaviour and therefore is not shown in the plot.

As the integer $n$ increases the energy correction continues to have a positive, but significantly smaller, imaginary part. For $n > 2$ the last term in Eq.~(\ref{1.3.33}) becomes dominant and the transmission is essentially controlled by flips into the continuum. 

\section{Concluding remarks}
\label{end}

In the present paper we have clearly demonstrated in a systematic way that a suitable reformulation of GPPA \cite{Diakonos2012} is an improved version of the standard adiabatic perturbation theory. The proposed improvement is based on the introduction of the loop amplitudes obtained by the resummation of all elementary transitions beginning and ending at the same state in the discrete part of the spectrum.  When applied to a bound system this technique expresses a general transition amplitude in terms of the loop contributions and the elementary transitions between strictly different states. In this form the expansion series of a transition amplitude terminates or converges. We have shown that the loop contributions may produce damping factors connected with the time life of a certain bound state. Using a specific example it is demonstrated that in a driven bound system such damping factors can be induced by a polychromatic driving term. It is also shown that in driven systems with mixed instantaneous energy spectrum the discrete part becomes necessarily quasi-bound, acquiring a finite life leading to a Fano resonance in the corresponding transmission profile. Thus, the  present work clearly establishes the reformulated GPPA as a valuable tool for gaining insight into fundamental properties of driven systems beyond that obtained by the usual perturbative methods.

\appendix
\section{}
\label{appen}

We will now show how the rearrangement of the GPPA terms can be achieved in practice. Initially we consider the first and the third terms in the series of Eq.~(\ref{1.1.9}), namely:

\begin{equation}
\label{A.1.1}
{X_{n0}^{\left( 1 \right)} = \frac{i}{\hbar }\int\limits_{{t_i}}^t {d{t_1}} {\varphi _{n0}}\left( {{t_1}} \right)}
\end{equation}

\noindent
and

\begin{equation}
\label{A.1.2}
{X_{n0}^{(3)} = {\left( {\frac{i}{\hbar }} \right)^3}\sum\limits_{{n_1},{n_2}} {\int\limits_{{t_i}}^t {d{t_3}\int\limits_{{t_i}}^{{t_3}} {d{t_2}} } \int\limits_{{t_i}}^{{t_2}} {d{t_1}{\varphi _{n{n_2}}}\left( {{t_3}} \right)} {\varphi _{{n_2}{n_1}}}\left( {{t_2}} \right){\varphi _{{n_1}0}}\left( {{t_1}} \right)}}
\end{equation}

Isolating the ${n_2} = 0$ contribution in Eq.~(\ref{A.1.2}) and combining it with 
Eq.~(\ref{A.1.1}) we rewrite the first order coefficient as:

\begin{equation}
\label{A.1.3}
{X_{n0}^{\left( 1 \right)} \to \frac{i}{\hbar }\int\limits_{{t_i}}^t {d{t_3}} {\varphi _{n0}}\left( {{t_3}} \right)\left[ {1 + {{\left( {\frac{i}{\hbar }} \right)}^2}\sum\limits_{{n_1}} {\int\limits_{{t_i}}^{{t_3}} {d{t_2}\int\limits_{{t_i}}^{{t_2}} {d{t_1}} {\varphi _{0{n_1}}}\left( {{t_2}} \right){\varphi _{{n_1}0}}\left( {{t_1}} \right)} } } \right]}
\end{equation}

\noindent
It is obvious that the inclusion of analogous terms occurring at higher orders permits the following redefinition of the first order contribution to the transition amplitude 
(\ref{1.1.9}):

\begin{equation}
\label{A.1.4}
{X_{n0}^{{{\left( 1 \right)}^\prime }} = \frac{i}{\hbar }\int\limits_{{t_i}}^t {d{t_1}} {\varphi _{n0}}\left( {{t_1}} \right){X_{00}}\left( {{t_1},{t_i}} \right)}
\end{equation}

\noindent
In a similar manner we deal with the second and the fourth terms in Eq.~(\ref{1.1.9}):

\begin{equation}
\label{A.1.5}
{X_{n0}^{\left( 2 \right)} = {\left( {\frac{i}{\hbar }} \right)^2}\sum\limits_{{n_1}} {\int\limits_{{t_i}}^t {d{t_2}} \int\limits_{{t_i}}^{{t_2}} {d{t_1}{\varphi _{n{n_1}}}\left( {{t_2}} \right)} {\varphi _{{n_1}0}}\left( {{t_1}} \right)}}
\end{equation}

\noindent
and

\begin{equation}
\label{A.1.6}
{X_{n0}^{(4)} = {\left( {\frac{i}{\hbar }} \right)^4}\sum\limits_{{n_1},{n_2},{n_3}} {\int\limits_{{t_i}}^t {d{t_4}...} \int\limits_{{t_i}}^{{t_2}} {d{t_1}{\varphi _{n{n_3}}}\left( {{t_4}} \right)} {\varphi _{{n_3}{n_2}}}\left( {{t_3}} \right){\varphi _{{n_2}{n_1}}}\left( {{t_2}} \right){\varphi _{{n_1}0}}\left( {{t_1}} \right)}}
\end{equation}

\noindent
Again, isolating the ${n_2} = 0$ contribution and combining it with Eq.~(\ref{A.1.5}) we get:

\begin{equation}
\label{A.1.7}
\begin{split}
X_{n0}^{\left( 2 \right)} \to & {\left( {\frac{i}{\hbar }} \right)^2}\sum\limits_{{n_3}} {\int\limits_{{t_i}}^t {d{t_4}} \int\limits_{{t_i}}^{{t_4}} {d{t_3}{\varphi _{n{n_3}}}\left( {{t_4}} \right)} {\varphi _{{n_3}0}}\left( {{t_3}} \right)}  \times \\ & \times \left[ {1 + {{\left( {\frac{i}{\hbar }} \right)}^2}\sum\limits_{{n_1}} {\int\limits_{{t_i}}^{{t_3}} {d{t_2}\int\limits_{{t_i}}^{{t_2}} {d{t_1}} {\varphi _{0{n_1}}}\left( {{t_2}} \right){\varphi _{{n_1}0}}\left( {{t_1}} \right)} } } \right]
\end{split}
\end{equation}

\noindent
and the inclusion of the higher order contributions leads to the following redefinition of Eq.~(\ref{A.1.5}):

\begin{equation}
\label{A.1.8}
{X_{n0}^{{{\left( 2 \right)}^\prime }} = {\left( {\frac{i}{\hbar }} \right)^2}\sum\limits_{{n_1}} {\int\limits_{{t_i}}^t {d{t_2}} \int\limits_{{t_i}}^{{t_2}} {d{t_1}{\varphi _{n{n_1}}}\left( {{t_2}} \right)} {\varphi _{{n_1}0}}\left( {{t_1}} \right){X_{00}}\left( {t,{t_i}} \right)}}
\end{equation}

\noindent
It is straightforward to extend this recombination scheme to all the terms appearing in the expansion (\ref{1.1.9}):

\begin{equation}
\label{A.1.9}
\begin{split}
{X_{n0}} = & \sum\limits_{r = 1}^\infty  {X{{_{n0}^{(r)}}^\prime }}  =  \\ & \sum\limits_{r = 1}^\infty  {\sum\limits_{{n_1} \ne 0,...,{n_{r - 1}} \ne 0} {{{\left( {\frac{i}{\hbar }} \right)}^r}\int\limits_{{t_i}}^t {d{t_r}} \int\limits_{{t_i}}^{{t_r}} {d{t_{r - 1}}...\int\limits_{{t_i}}^{{t_2}} {d{t_1}} } {\varphi _{n{n_{r - 1}}}}\left( {{t_r}} \right) \times}} \\ &\times {{ {\varphi _{{n_{r - 1}}{n_{r - 2}}}}\left( {{t_{r - 1}}} \right)...{\varphi _{{n_1}0}}\left( {{t_1}} \right)} {X_{00}}} \left( {{t_1},{t_i}} \right)
\end{split}
\end{equation}

\noindent
This rearrangement can be further extended reconsidering the first and the third order terms:

\begin{equation}
\label{A.1.10}
{X_{n0}^{{{\left( 1 \right)}^\prime }} = \frac{i}{\hbar }\int\limits_{{t_i}}^t {d{t_1}} {\varphi _{n0}}\left( {{t_1}} \right){X_{00}}\left( {{t_1},{t_i}} \right)}
\end{equation}

\noindent
and

\begin{equation}
\label{A.1.11}
\begin{split}
X_{n0}^{{{\left( 3 \right)}^\prime }} = & {\left( {\frac{i}{\hbar }} \right)^3}\sum\limits_{{n_1} \ne 0,{n_2} \ne 0} {\int\limits_{{t_i}}^t {d{t_3}} \int\limits_{{t_i}}^{{t_3}} {d{t_2}} \int\limits_{{t_i}}^{{t_2}} {d{t_1}{\varphi _{n{n_2}}}\left( {{t_3}} \right)} {\varphi _{{n_2}{n_1}}}\left( {{t_2}} \right){\varphi _{{n_1}0}}\left( {{t_1}} \right) \times} \\ & {\times {X_{00}}\left( {{t_1},{t_i}} \right)}.
\end{split}
\end{equation}

\noindent
From the last expression we isolate the  ${n_1} = n$ term and we recombine it with 
(\ref{A.1.10}):

\begin{equation}
\label{A.1.12}
\begin{split}
X_{n0}^{{{\left( 1 \right)}^\prime }} \to & \frac{i}{\hbar }\int\limits_{{t_i}}^t {d{t_1}} \left[ {1 + \sum\limits_{{n_2} \ne 0} {{{\left( {\frac{i}{\hbar }} \right)}^2}\int\limits_{{t_1}}^t {d{t_3}\int\limits_{{t_1}}^{{t_3}} {d{t_2}{\varphi _{n{n_2}}}\left( {{t_3}} \right){\varphi _{{n_2}n}}\left( {{t_2}} \right)} } } } \right] \times \\ & \times {\varphi _{n0}}({t_1}){X_{00}}\left( {{t_1},{t_i}} \right)
\end{split}
\end{equation}

\noindent
Including all the higher order contributions we define the following "renormalized" first order coefficient:

\begin{equation}
\label{A.1.13}
{X_{n0}^{\left( 1 \right)R} = \frac{i}{\hbar }\int\limits_{{t_i}}^t {d{t_1}} X_{nn}^{\left[ 0 \right]}\left( {t,{t_1}} \right){\varphi _{n0}}\left( {{t_1}} \right){X_{00}}\left( {{t_1},{t_i}} \right)}
\end{equation}

\noindent
Working along the same lines we renormalize all of the terms appearing in the expansion (\ref{A.1.9}) and we obtain Eq.~(\ref{1.1.10}).

\end{document}